\input{aipcheck}

\documentclass[
    ,final            
  ]
  {aipproc}

\layoutstyle{6x9}

\begin{document}

\title{Galactic Twins of the Nebula Around SN 1987A: Hints that LBVs
  may be supernova progenitors}


\author{Nathan Smith}{
  address={Astronomy Department, University of California, 601
  Campbell Hall, Berkeley CA 94720} }

\begin{abstract}

I discuss outstanding questions about the formation of the ring nebula
around SN1987A and some implications of similar ring nebulae around
Galactic B supergiants.  There are notable obstacles for the formation
of SN1987A's bipolar nebula through interacting winds in a transition
from a red supergiant to a blue supergiant.  Instead, several clues
hint that the nebula may have been ejected in an LBV-like event.  In
addition to the previously known example of Sher~25, there are two
newly-discovered Galactic analogs of SN1987A's ringed nebula.  Of
these three Galactic analogs around blue supergiants, two (Sher~25 and
SBW1) have chemical abundances indicating that they have not been
through a red supergiant phase, and the remaining ringed bipolar
nebula surrounds a luminous blue variable (HD168625).  Although
SK$-$69~202's initial mass of $\sim$20 M$_{\odot}$ is lower than those
atributed to most LBVs, it is not far off, and the low-luminosity end
of the LBV phenomenon is not well defined.  Furthermore, HD168625's
luminosity indicates an initial mass of only $\sim$25 M$_{\odot}$,
that of SBW1 is consistent with $\sim$20 M$_{\odot}$, and there is a
B[e] star in the SMC with an initial mass of $\sim$20 M$_{\odot}$ that
experienced an LBV outburst in the 1990s.  These similarities may be
giving us important clues about Sk$-$69~202's pre-SN evolution and the
formation mechanism of its nebula.

\end{abstract}
\maketitle

\section{INTRODUCTION}

It is commonly assumed that bipolar nebulae around massive stars
consist of slow ambient material ejected as a red supergiant that is
swept-up by the faster wind of a hot supergiant.  To create the
bipolar shape, the surrounding slow wind must have an equatorial
density enhancement (i.e. a disk); the consequent mass loading near
the equator slows the expansion and gives rise to a pinched waist and
bipolar structure.  However, it is unclear how the required
pre-existing disk could have been formed. One does not normally expect
RSG or AGB stars to rotate rapidly, so a disk-shedding scenario
probably requires the tidal influence of a companion star during prior
evolutionary phases in order to add sufficient angular momentum. In
the case of SN~1987A, a binary merger would be required for this
particular scenario to work (Collins et al.\ 1999; Podsiadlowski,
these proceedings).  However, there are reasons to question a binary
merger scenario for the formation of SN1987A's nebula:

1.  A merger model followed by a transition from a RSG to BSG requires
that these two events be synchronized with the supernova event itself,
requiring that the best observed supernova in history also happens to
be a very rare event.  One could easily argue, though, that the merger
and the blue loop scenario would not need to have been invented if
SN1987A had occurred in a more distant galaxy where it would not have
been so well-observed (i.e. we wouldn't know about the bipolar nebula
or its BSG progenitor).  Admittedly, this is a bit of a
``faith-based'' argument.

2.  After the RSG swallowed a companion star and then contracted to
become a BSG, it should have been rotating at or near its critical
breakup velocity.  Even though pre-explosion spectra (Walborn et al.\
1989) do not have sufficient resolution to measure line profiles,
Sk--69$^{\circ}$202 showed no evidence of rapid rotation (e.g., like a
B[e] star spectrum).  Instead, Sk--69$^{\circ}$202 had the spectrum of
an entirely normal B3 supergiant.

3.  Particularly troublesome is that this merger and RSG/BSG
transition would need to occur twice. From an analysis of light echoes
for up to 16 yr after the supernova, Sugerman et al.\ (2005; in
addition, see the contribution by Arlin Crotts in these proceedings)
have identified a much larger bipolar nebula with the {\it same axis
orientation} as the more famous inner triple ring nebula.  If a merger
and RSG/BSG transition are to blame for the bipolarity in the
triple-ring nebula, then what caused it in the older one?

\section{LBV\lowercase{s} AS SUPERNOVA PROGENITORS}

Perhaps a more natural explanation would be that Sk--69$^{\circ}$202
suffered a few episodic mass ejections analogous to luminous blue
variable (LBV) eruptions in its BSG phase (see Smith 2007).  There is
mounting evidence that LBVs do, in fact, explode as supernovae (see
Gal-Yam et al.\ 2007; Smith et al.\ 2007b; Smith 2007; Smith \& Owocki
2006; Kotak \& Vink 2006).  We all know this is not supposed to
happen, because we expect very massive stars to shed their hydrogen
envelopes and live on for another few hundred thousand years as
Wolf-Rayet stars after the LBV phase.  The empirical evidence that
some massive stars seem to die prematurely as LBVs therefore presents
a challenge to current evolution models.  It also highlights our poor
understanding of the LBV phase, since these stars, at least the more
massive ones, are supposed to be in transition from the end of core-H
burning to core-He burning.

We normally think of the LBVs as a late stage of evolution for very
massive stars with initial masses above 40--50 M$_{\odot}$ on their
way to becoming Wolf-Rayet stars.  However, there is also a
lower-luminosity group of LBVs exhibiting similar behavior that fall
in a range of luminosities corresponding to initial masses of 25--35
M$_{\odot}$ (see Smith, Vink, \& de Koter 2004).  These
lower-luminosity LBVs are generally assumed to be post-RSGs in order
that RSG mass loss has lowered their M/L ratio enough to make them
susceptible to the LBV instability.  Although they are not normally
discussed as potential SN progenitors in the literature, we should
expect them to be --- as post-RSGs, these lower-luminosity LBVs likely
represent their final evolutionary state (i.e. they will not become WR
stars, because empirical studies show that all WR stars come from
stars with initial masses above 30 M$_{\odot}$, and most above 50
M$_{\odot}$; Humphreys, Nichols, \& Massey 1985).

The lower luminosity boundary of this LBV group is not clearly
established by observations.  Sk--69~202 is thought to have had an
initial mass close to 20 M$_{\odot}$, and at first glance this seems
too low to allow it to be a normal LBV.  However, the B[e] star R4 in
the Small Magellanic Cloud may offer a precedent at the same
luminosity as the progenitor of SN1987A; R4 is consistent with a 20
M$_{\odot}$ evolutionary track, and it experienced an LBV outburst in
the late 1980's (Zickgraf et al.\ 1996).  R4 also has elevated
nitrogen abundances comparable to the nebula around SN~1987A, so it is
also a post-RSG star.  As we will see below, two other ringed blue
supergiant stars have similar low initial masses of 20 to 25
M$_{\odot}$.

\begin{figure}
  \includegraphics[height=.5\textheight]{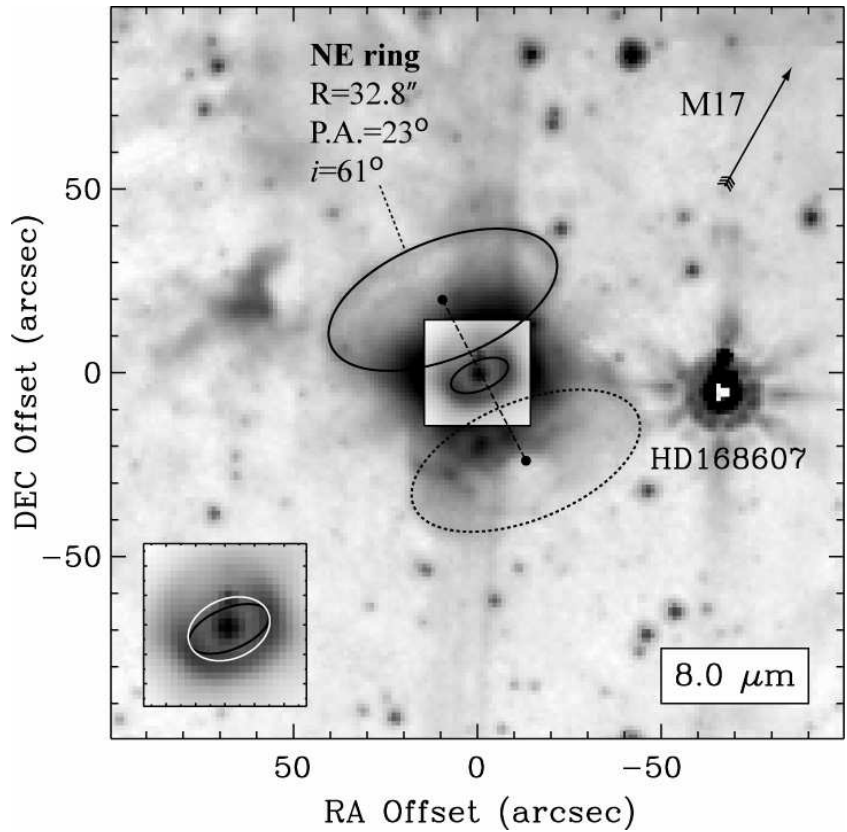} \caption{An 8 $\mu$m
  Spitzer/IRAC image of the LBV candidate HD168625 from Smith (2007).
  It shows a nebula with a geometry very much like that around
  SN1987A, but in this case the bipolar shape probably originated
  during the ejection by the central LBV star and not from interacting
  winds.}
\end{figure}

\section{GALACTIC ANALOGS OF SN1987A'\lowercase{s} NEBULA}

We can gain further insight to the formation of SN1987A's ring nebula
and its pre-SN evolutionary state by studying analogs of it around
massive stars in our own Galaxy and asking what those stars are like.
Are they post-merger products?  Close binaries?  Rapid rotators?  Do
their chemical abundances indicate post-RSG evolution? Three close
analogs in the Milky Way are currently known:

{\bf Sher 25 in NGC3603:} HST images of this B1.5 supergiant (not
shown here) reveal a remarkable equatorial ring with the same radius
as the one around SN1987A, plus bipolar ejecta (Brandner et al.\
1997).  Although the nebula has moderate N-enrichment, Smartt et al.\
(2002) find that the N abundance is too low to be the result of
post-RSG evolution.  In fact, the stellar luminosity is above the
limit where no RSGs are seen.  Thus, the nebula around Sher 25
probably did not form from interacting winds during a RSG-BSG
transition.  The star's spectrum shows fairly narrow lines (Smartt et
al.\ 2002) and there is not indication that it is an extremely rapid
rotator (we know sin $i$, presumably, from the tilt angle of the
nebular ring).  It is also not yt known to be a close binary --
although if it is a binary, that obviously rules out the merger
hypothesis because the ring has already formed.

{\bf HD168625 near M17:} This LBV candidate has a luminosity much
closer to the progenitor of SN1987A than Sher 25, consistent with an
initial mass of $\sim$25 M$_{\odot}$.  Its nebula has an equatorial
ring, and it is the only object known so far to also show polar rings
like SN1987A (Fig.\ 1; see Smith 2007).  This nebula makes it our
Galaxy's closest analog to the one around the progenitor of SN1987A.
Its LBV status is interesting, since LBVs are known to have eruptive
episodes of high mass loss (e.g., Smith \& Owocki 2006) and are often
surrounded by bipolar nebulae.  Based on various observed properties
of the nebula, I have argued (Smith 2007) that the nebula was probaby
ejected as an LBV rather than ejected as a RSG and shaped afterward by
a fast BSG wind.  The central sttar has been studied extensively in
order to study its possible LBV-like variability (e.g., Chentsov \&
Gorda 2004).  It is not a rapid rotator and is not known to be a
binary.  Its nebula may be moderately enhanced with CNO products,
although uncertainties in the N abundance (Nota et al.\ 1996) make it
difficult to determine if it really is a post-RSG.

{\bf SBW1 in the Carina Nebula:} This equatorial ring nebula (Fig.\ 2)
also has the same 0.2~pc radius as the one around SN1987A, and the
central B1.5 supergiant has essentially the same luminosity as
Sk-69$^{\circ}$202, consistent with an initial mass of roughly 20
M$_{\odot}$.  The age and expansion speed of the ring around SBW1 are
within a factor of 2 of the equatorial ring of SN1987A.  It is seen
toward the Carina Nebula, but it is probably more distant, at
$\sim$7kpc (Smith et al.\ 2007a).  Its nebula shows no evidence for
N-enrichment; the N abundance is roughly solar (Smith et al\ 2007a).
Thus, this ring formed even though the star has never been a RSG.  The
central star is not an extremely rapid rotator either (Smith et al.\
2007a).

\begin{figure}
  \includegraphics[height=.45\textheight]{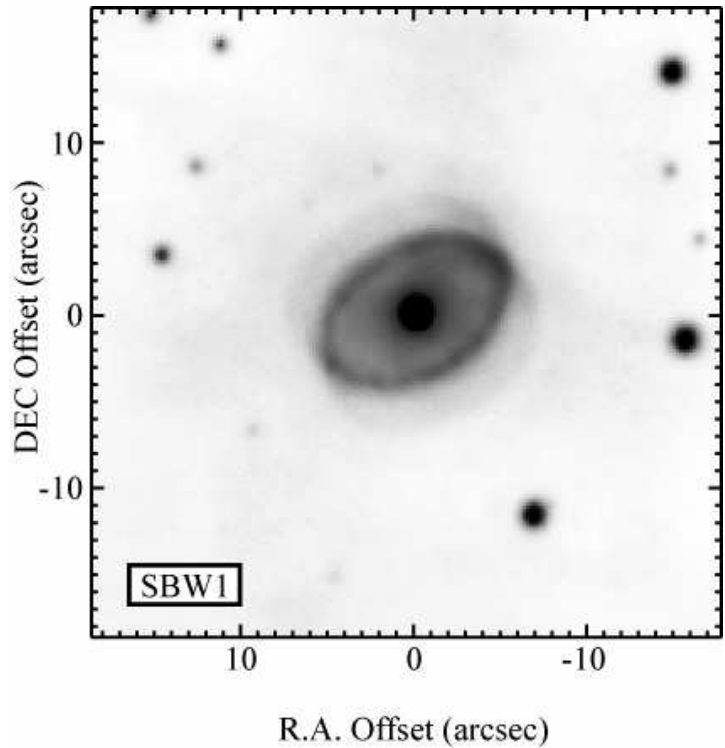} \caption{An H$\alpha$
  image of the ring nebula SBW1 in the Carina Nebula from Smith et
  al.\ (2007a), surrounding a B1.5 Iab supergiant.  It has the same
  radius (0.2 pc) as the ring around SN1987A and the star has the same
  luminosity as the progenitor of SN1987A, but it has solar N
  abunance, indicating that it has not yet been a RSG.  In addition to
  the equatorial ring, it appears to have faint bipolar lobes as well.}
\end{figure}

Of the three examples of ring nebulae around BSGs that are our
Galaxy's closest known analogs to the nebula around the progenitor of
SN1987A, two could not have been red supergiants because of their
chemical abundances, and one was ejected as an LBV.  Thus, of the
three examples known, {\it none} were formed by interacting winds
during a RSG to BSG transition.  This proves that there must be some
other physical mechanism that can eject equatorial rings and bipolar
nebulae.  The best candidate is an intrinsically bipolar ejection by a
rotating LBV, or an episodic mass ejection analogous to LBV outbursts.
The star does not necessarily need to have a high angular velocity, as
the effects of rotational shaping can be enhanced in a star with even
moderate angular speed if it is near the Eddington limit. (During LBV
eruptions, the star is thought to increase its bolometric luminosity
and approach or even exceed the Eddington limit.)  This also hints
that SN1987A and other type II SNe with circumstellar material did not
necessarily transition recently from the RSG phase; instead, they may
have been in an LBV-like phase before explosion.  If LBVs can be SN
progenitors, it puts a rather embarassing spotlight on our current
lack of an explanation for the LBV instability.

\section{FORMATION OF THE TRIPLE RINGS?}

So, the obvious question then is whether or not a single rotating star
is able to produce a triple-ring nebula like SN1987A's during an
LBV-like event.  I mean, come on -- how can a star do that?  This
challenge for a single star model is all the more daunting considering
how well Phil Podsiadlowski's merger model can explain the detailed
structure of the nebula (see Podsiadlowski, these proceedings; Morris
\& Podsiadlowski 2007).  I was quite impressed by that, but I'm not
ready to throw in the towel just yet.

As I noted earlier, the key obstacles for the binary
merger/wind-interaction model are that it requires the nearest SN in
400 years to be nearly synchronized with an extremely unlikely event,
it requires that the merger product was an extremely rapid rotator
contrary to observations, and it needs to have happened at least twice
(!) because the outer nebula has the same bipolar axis.  These three
obstacles are eliminated for the LBV-ejection hypothesis because the
timescale between successive LBV eruptions of a few thousand years is
comparable to the dynamical time of the nebula (no longer
synchronized), extremely rapid rotation (i.e. high angular velocity)
is not needed to achieve critical rotation because of the reduced
$g_{\rm eff}$ near the Eddington limit, and we know that LBV eruptions
happen repeatedly with the same bipolar geometry (the best example of
that is $\eta$ Car; see Smith 2005).

I should be clear that in this scenario, Sk--69 202 was still in a
post-RSG phase as required by its chemical abundances, but it did not
need to explode so soon after making the RSG-BSG transition.  It could
have made that transition and then could have been in an LBV phase for
$\sim$10$^5$ years, during which time it suffered a few major LBV-like
mass ejection episodes separated by thousands of years before finally
exploding.

Now, I'll admit that I'm still at a bit of a loss as to how one might
form the triple rings in this scenario. (However, keep in mind that
the LBV star HD168625 was apparently able to form equatorial and polar
rings, and its central star is not a rapidly-rotating post-merger
product and is not known to be a binary despite attempts to monitor
its spectral variability.)  That seems like a difficult hydro problem
connected to the mass ejection mechanism itself (see below).  However,
there is a relatively simple explanation for how a single rotating
star might produce an equatorial disk/ring {\it and} bipolar lobes
simultaneously.  Tis is a logical first step, or a prerequisite,
toward producing an equatorial ring with a pair of polar rings in a
single-star scenario.  Suppose that a rotating star increases its
luminosity during an LBV eruption, and reaches near-critical rotation
while ejecting matter from its surface.  That star will be oblate and
will likely have substantial gravity darkening (e.g., von Zeipel
1924), leading to a faster and denser polar wind (e.g., Owocki et al.\
1996).  For a short duration ejection, this will naturally lead to a
pair of hollow bipolar lobes after expansion to large radii.  There is
a competing effect that will also enhance the density at the equator.
At latitudes near the equator where ejection speeds are low or
comparable to the rotational speed, centrifugal forces will divert
ballistic trajectories toward the equatorial plane where material can
collide with ejecta from the opposite hemisphere to form a disk
(analogous to the wind-compressed disk model of Bjorkman
\& Cassinelli 1993).  Combined, these two effects can lead to bipolar
+ equatorial ejection from a rotating star.  There is insufficient
space to present and defend the model with appropriate detail here
(and I didn't do so in my talk), so I'll just point the reader to
Smith \& Townsend (2007) for the details of how this may work.

\begin{figure}
  \includegraphics[height=.3\textheight]{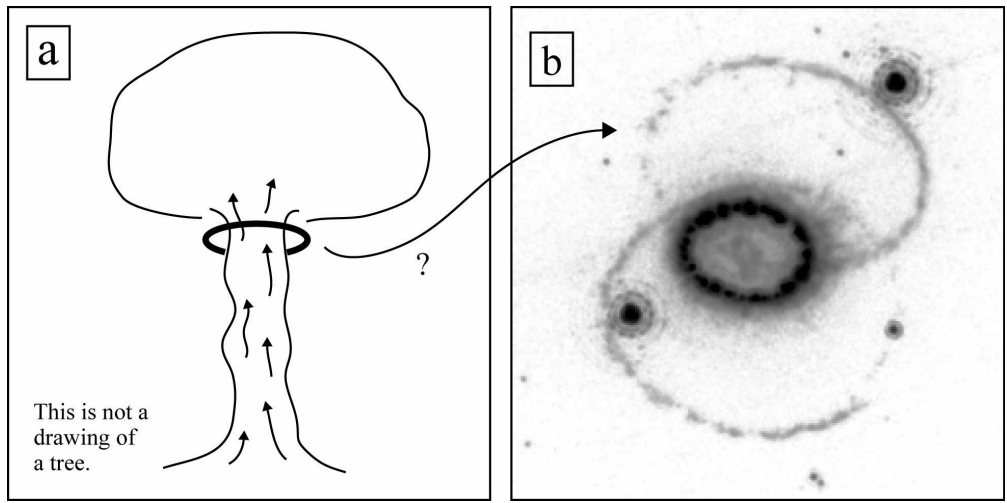}\caption{The cartoon in
  Panel a shows an unfortunate hydrodynamic situation that forms a
  vortical ring, with possible relevance to the polar rings of
  SN1987A's nebula shown in Panel b.  The text gives a more detailed
  explanation.}
\end{figure}

Since this contribution does not need to pass the muster of a referee,
I'll take this opportunity to go a step further and discuss a
half-baked idea that I mentioned to a few people at the conference as
a potential explanation for how one might get polar rings (the
equatorial ring around SN1987A is easy to explain, by comparison).  I
can think of one possibly relevant analogy of a hydrodynamic situation
that leads to the formation of thin rings --- an atomic bomb explosion
rising in the Earth's atmosphere.  In footage of such explosions (see
Kubrick 1964), one sometimes sees a ring forming around the stem of
the mushroom cloud as it rises.  This is a vortical ring formed from
shear between the surrounding gas and the rising hot plume in the stem
of the mushroom cloud.  There may be an astrophysical application of
this: in a non-spherical surface explosion from a near-critically
rotating star, the poles of the star will be hotter and the escape
velocity (and hence the ejecta speed) will be faster, as noted
earlier.  Moving from equator to pole, then, there will be a gradient
in the expansion speed and one could imagine that shear might occur at
some mid-latitude during the explosion, in a manner analogous to a
mushroom cloud or smoke rings.  Reflected around the rotation axis at
mid latitudes both above and below the equator, this might lead to a
pair of polar rings if the density structure is frozen-in to the
expanding ejecta.  Testing the plausibility of this idea will require
a detailed numerical simulation, of course, because currently it is
little more than a suspicion of mine that someting like this might
work.  In any case, the rings around SN1987A and HD168625, as well as
vaguely similar structures around bipolar planetary nebulae (e.g.,
Balick \& Frank 2002), are still persistent unsolved astrophysical
problems that at the same time hold critical clues to the nature of
the central stars.



\begin{theacknowledgments}

I would like to thank the conference organizers for partial financial
support and for organizing an enjoyable and productive meeting.

\end{theacknowledgments}



\end{document}